\documentclass[12pt]{JHEP3}

\usepackage{epsfig}
\usepackage{epstopdf}
\epsfclipon
\usepackage{multicol}
\usepackage{epsfig,bm}
\usepackage{amssymb,amsmath}

\newcommand{\roughly}[1]{\mathrel{\raise.3ex\hbox{$#1$\kern-0.85em
\lower1ex\hbox{$\sim$}}}}

\def\Expect#1{\langle #1 \rangle}

\def\be{\begin{equation}}
\def\beq\begin{equation}
\def\ee{\end{equation}}
\def\bea{\begin{eqnarray}}
\def\eea{\end{eqnarray}}

\def\endignore{}
\def\ignore #1\endignore{} % use to "comment out" text

\def\nn{\nonumber}

\def\pref#1{(\ref{#1})}

\def\beq{\begin{equation}}
\def\eeq{\end{equation}}
\def\beqa{\begin{eqnarray}}
\def\eeqa{\end{eqnarray}}

\def\cH{{\cal H}}

\def\cO{{\cal O}}

\def\ssA{{\scriptscriptstyle A}}

\def\ssC{{\scriptscriptstyle C}}

\def\ssN{{\scriptscriptstyle N}}

\def\ssR{{\scriptscriptstyle R}}
\def\ssT{{\scriptscriptstyle T}}

\def\UV{{\scriptscriptstyle U \kern-.1emV}}
\def\IR{{\scriptscriptstyle I\kern-.1em R}}

\newcommand{\bmat}{\left(\begin{array}}
\newcommand{\emat}{\end{array}\right)}

\def\-{\hphantom{-}}

\def\s2{\frac{1}{2}}

\def\IF{\relax{\rm I\kern-.18em F}}
\def\II{\relax{\rm I\kern-.18em I}}
\def\IP{\relax{\rm I\kern-.18em P}}
\def\IC{\relax{\rm I\kern-.48em C}}
\def\IK{\relax{\rm I\kern-.20em K}}
\def\IM{\relax{\rm I\kern-.25em M}}

\def\Dsl{\,\raise.15ex\hbox{/}\mkern-13.5mu D} %this one can be subscripted
\def \one{\relax{\rm 1\kern-.26em I}}

\def\exd{{\rm d}}

\title{Super-Hubble de Sitter Fluctuations \\ and the Dynamical RG}

\author{C.P.~Burgess,${}^{1,2}$ L. Leblond,${}^{2}$
 R. Holman${}^{3}$ and S. Shandera${}^{2}$\\
$^1$ Department of Physics and Astronomy,
  McMaster University, \\ \qquad \qquad Hamilton, Ontario, Canada;\\
$^2$ Perimeter Institute for Theoretical Physics,
  Waterloo, Ontario, Canada;\\
$^3$ Department of Physics, Carnegie Mellon University,\\
 \qquad \qquad Pittsburgh, Pennsylvania 15213;\\
 }

\date{}

\abstract{Perturbative corrections to correlation functions for
interacting theories in de Sitter spacetime often grow secularly
with time, due to the properties of fluctuations on super-Hubble
scales. This growth can lead to a breakdown of perturbation theory
at late times. We argue that Dynamical Renormalization Group (DRG)
techniques provide a convenient framework for interpreting and
resumming these secularly growing terms. In the case of a massless
scalar field in de Sitter with quartic self-interaction, the
resummed result is also less singular in the infrared, in
precisely the manner expected if a dynamical mass is generated.
We compare this improved infrared behavior with large-$N$ expansions when
applicable.}

\preprint{}


\begin{document}

%\newpage

%===================================================================================

\section{Introduction}

Detailed observations of the properties of the Cosmic Microwave
Background (CMB) \cite{CMBobs} have lifted cosmology to a
precision science, raising the bar for theorists who compute the
predictions with which these observations must be compared in
order to extract their meaning. This is true in particular for
calculations of primordial fluctuations from very early
inflationary environments, spurring detailed studies of various
kinds of corrections to the early classic computations
\cite{CMBth1}. Such studies \cite{Sloth}---\cite{JMPW} are forcing
a re-examination of how to interpret the relatively unusual
properties of fluctuations over super-Hubble distances that have
long been known to arise when computing for de Sitter, and near-de
Sitter, spacetimes \cite{CMBthn}.

Super-Hubble fluctuations raise two, related, puzzles for
perturbation theory in de Sitter space. One of these is the
presence of infrared divergences, and the other is the appearance
of secular time dependence in successive orders of perturbation
theory. In this paper we focus on the secular growth, which is
troublesome because it undermines the validity of perturbative
calculations at late times, and so leaves open the ultimate
late-time fate of the system. We argue that this late-time
behavior can be understood in a controlled way by using the {\it
Dynamical Renormalization Group} (DRG) \cite{DRG} --- a technique
for dealing with similar problems in condensed matter physics.
(See ref.~\cite{DRGcosmo} for some earlier applications of DRG
methods to cosmology.) The DRG is designed for the situation where
perturbative solutions to time-dependent equations are limited by
the appearance of a secular and growing time-dependence that
restricts the domain of validity of perturbative methods. The DRG
allows the late-time behavior to be inferred nonetheless, by using
renormalization group (RG) methods to extend the domain of
validity of the perturbative solution. This technique is
well-adapted to perturbative calculations in de Sitter spacetimes,
which often reveal such a secular dependence on the cosmological
scale-factor, $a(t)$.

We believe it is important to distinguish RG methods, such as
these, from wholesale resummation techniques that identify and
resum specific infinite classes of graphs (such as chain, planar,
daisy or cactus graphs). Although RG methods can also be
interpreted as resummations, these need not be explicitly
performed graphically since the leading logarithms can instead be
obtained by integrating the appropriate RG equation. Explicit
resummations, such as arise for hard thermal loops or in large-$N$
expansions for example, are normally required to describe a more
serious breakdown of perturbation theory than is necessary simply
to track large logarithms.

Although our technique is aimed at understanding the late-time
behavior of de Sitter correlation functions, we find it also sheds
light on the related infrared divergences in de Sitter space. This
is because the resummed late-time correlations can also be less
singular in the infrared, and this improved IR behavior can cure
the infrared singularity. To further explore this connection, we
examine the special case of $N$ scalars and compare the DRG
resummed results with those obtained using large-$N$ techniques.
This comparison shows how the infrared behavior identified by the
DRG properly captures the effects of dynamical mass generation in
the large-$N$ theory, which is also responsible for removing the
IR divergences. The DRG resummed result resembles a mass even when
$N=1$, however, and this can be interpreted as evidence in favour
of stochastic arguments for dynamical mass generation for the
long-wavelength de Sitter modes \cite{Starobinsky}. However, when
we perform the DRG analysis for a cubic scalar interaction, we
instead find a substantially different result that does not have a
similar interpretation in terms of a mass. We argue that this is
consistent with the expectation that a dynamical mass is not
generated in this particular case.

It is important to distinguish between two different types of
possible secular growth that can appear in de Sitter calculations.
The first kind counts the number of $e$-folds between horizon
crossing for a given mode with comoving momenta $k$ and some
conformal time $\tau$, generically of the form $\ln(-k\tau)$. The
second kind counts the number of $e$-folds between some fixed time
$\tau_0$, such as the beginning of inflation, and the time of
interest \cite{preHIosc}. For interacting
theories the first of these can arise at next-to-leading order even in de Sitter
invariant situations. The second, by contrast, arises in
situations when de Sitter invariance is broken by the existence of
a preferred time scale, and can appear even at leading order
without self-interactions. While in principle, the DRG could be
applied to both cases, in this paper we concentrate only on the de
Sitter invariant case, tracking loop-induced secular growth from
horizon crossing.

Large logs of the form $\ln(-k\tau)$ have been dealt with in
various ways in the literature, mostly in the context of
observable curvature fluctuations after a period of inflation. In
that case, one possibility is to simply keep them in the
calculation of the scalar field fluctuations. Indeed for the modes
of observational interest this factor is at most of order $60$
which, while large, may not be so large as to destabilize the
perturbation series for practical purposes. More importantly, for
a finite period of inflation the real variable of interest is the
curvature perturbation, which is conserved outside the horizon for
single-field scenarios. In that case one may use the quantum
calculation to evaluate correlations of scalar field fluctuations
near horizon crossing (where the log piece is negligible), then
change gauge and follow the {\it classical} evolution of the
curvature perturbation between horizon exit and re-entry
\cite{Seery:2007wf}. In addition, van der Meulen and Smit
\cite{vMS} have argued more formally that the classical equation
of motion captures the leading log behavior, up to 1-loop. In this
work, we use the presence of the $\ln(-k\tau)$ terms as a tool for
understanding the late time behavior of fields in de Sitter space.
We believe this approach clarifies the physical picture in a way
that can ultimately be useful for observational questions as well.

We organize our presentation in the following way. First, \S2\
describes a representative one-loop calculation for a real scalar
field in de Sitter space, illustrating the infrared divergences
that emerge. This is followed, in \S3, by a quick review of both
traditional and dynamical RG methods, together with their
implication for the scalar field example of \S2. This calculation
shows in particular how the late-time behavior also resums the
large IR logs. \S4 then examines the case of $N$ scalar fields,
where it is shown that the DRG reproduces previous results for the
late-time limit of large-$N$ scalar fluctuations. \S5 then
performs a similar calculation for scalars having cubic
interactions. We discuss our conclusions in \S6. The appendices
contain a brief review of how the various couplings we examine are
renormalized.

\section{Next-to-Leading Order de Sitter calculations}

In this section we use a simple example to display the large
logarithms whose interpretation is the focus of this paper.

\subsection{Scalars in de Sitter space}

To this end consider a single real scalar field, $\phi$, coupled
to gravity through the action
\be
 S = - \int \exd^4x \; \sqrt{-g} \; \left[ \frac12 \,
 g^{\mu\nu} \, \partial_\mu \phi \, \partial_\nu \phi + V_0 +
 \frac{m_0^2}{2} \, \phi^2 + \frac{\lambda_0}{4!} \, \phi^4
 + \frac{\xi_0}{2} \, R \, \phi^2 \right] \,.
\ee
The subscript `$0$' is meant to remind that what appears here are
bare quantities, $m^2_0 = m_\ssR^2 + \delta m^2$, that include the
counter-terms required for the renormalization of ultraviolet (UV)
divergences. From this point on, we often drop the subscripts
`$R$' and `$0$,' except when emphasizing the renormalization, and
usually quantities without a subscript can be assumed to be
renormalized.

Our interest in what follows is in fluctuations about de Sitter
space, in which case $m$ and $\xi$ only appear through the
effective-mass combination\footnote{Our metric is mostly plus and
we use Weinberg's metric conventions \cite{WbgBook}, which differ
from those of MTW \cite{MTW} only by an overall sign in the
definition of the Riemann tensor. Consequently for 4D de Sitter
$R_{pmqn} = H^2 (g_{pn}g_{mq}-g_{pq}g_{mn})$, $R_{mn} = -3H^2
g_{mn}$ and $R = -12 H^2$. With these conventions a conformal
scalar has $\xi = - \frac16$. When necessary, we write the de
Sitter geometry using spatially flat coordinates and conformal
time, $\exd s^2 = a^2(\tau) (- \exd \tau^2 + \exd x_i \exd x^i)$
with $a(\tau) = -1/(\tau H)$ and $H$ constant.
%This system of coordinates only covers the upper half
%of the hyperboloid with $\tau$ ranging from $-\infty$
%to 0 while $x_i$ are valued in between $\pm \infty$.
}
\be
 M^2 = m^2 + \xi \, R
 = m^2 - 12 \, \xi H^2 \,.
\ee
The massless limit described in subsequent sections describes the
case $|M^2| \ll H^2$, which includes (but is not restricted to)
the case $\xi = m^2 = 0$. Notice that this region is {\em not} an
IR attractor of the renormalization group equations for the
constants $m^2$ and $\xi$, which (see Appendix A) over scales $\mu
\gg H$ renormalize according to
\be
 \mu \frac{\partial m^2}{\partial \mu} =
 \frac{\lambda \, m^2 }{16 \pi^2} \quad \hbox{and} \quad
 \mu \frac{\partial \xi}{\partial \mu} =
 \frac{\lambda }{16 \pi^2} \left(\xi + \frac16
 \right)   \,.
\ee
The solutions to these equations are attracted to the case of a
conformal scalar --- $m^2 = \xi + \frac16 = 0$, or $M^2 \to 2H^2$
--- as $\mu$ shrinks (provided $\mu$ remains larger than $H$).
When studying the massless case we must imagine these couplings to
lie on an RG trajectory that does not get to the attractor before
$\mu$ falls below $H$.

Scalar fluctuations within this theory may be computed using the
`in-in' formalism, which aims to evaluate the matrix element of an
operator,
\be
 \langle \cO(t) \rangle =
 \left\langle \hbox{in} \left| \left[ \overline T \exp \left(
 i \int_{t_{\rm in}}^t \exd t' \, \cH(t') \right) \right] \cO(t)
 \left[ T \exp \left( -i \int_{t_{\rm in}}^t \exd t' \, \cH(t')
 \right) \right] \right| \hbox{in} \right\rangle \,,
\ee
between two `in' vacuum states. To write this in the path integral
form, we double the number of fields, $\phi \to \{ \phi^+ , \phi^-
\}$, with one describing the evolution from $t_{\rm in}$ to $t$,
and the other giving the evolution back to $t_{\rm in}$ from $t$.

The corresponding correlation functions are
\bea
 G^{-+}(x,y) &=& i \langle \phi(x) \, \phi(y) \rangle \,,
 \qquad
 G^{+-}(x,y) = i \langle \phi(y) \, \phi(x) \rangle \,, \nn\\
 G^{++}(x,y) &=& \theta(x^0 - y^0) G^{-+}(x,y)
 + \theta(y^0 - x^0) G^{+-}(x,y) \,, \\
 G^{--}(x,y) &=& \theta(x^0 - y^0) G^{+-}(x,y)
 + \theta(y^0 - x^0) G^{-+}(x,y) \,.\nn
\eea
The identity $G^{++} + G^{--} = G^{+-} + G^{-+}$ ensures that only
three of these are independent. Defining $\phi_\ssC = \frac12(
\phi^+ + \phi^-)$ and $\phi_\Delta = \phi^+ - \phi^-$, this
constraint becomes $\langle \phi_\Delta
\phi_\Delta \rangle = 0$, and in the $\{ \phi_\ssC, \phi_\Delta
\}$ basis the correlation functions are conveniently written
\be
 \left(\begin{array}{cc}
  iG_\ssC & G_\ssR \\
  G_\ssA & 0 \\
 \end{array}\right)
 = W \left( \begin{array}{cc}
  G^{++} & G^{+-} \\
  G^{-+} & G^{--} \\
 \end{array} \right) W^\ssT \,,
\ee
where
\be
 W = \left( \begin{array}{rrr}
  \frac12 && \frac12 \\
  1 && -1 \\
 \end{array} \right) \,.
  \label{Wmatrix}
\ee
The advanced and retarded propagators are related by $G_\ssA(x,y)
= G_\ssR(y,x)$ and vanish in the coincidence limit. The free
propagators are determined by solving for the modes $u_{{\bf
k}}(\tau)$ of the field in this background and using
\be
 \phi(\tau,{\bf x}) = \int\frac{\exd^3k}{(2\pi)^3}
 \Bigl( e^{i {\bf k}\cdot {\bf x}} \alpha_{{\bf k}}
 u_{ {\bf k}}(\tau)+e^{-i {\bf k} \cdot {\bf x}}
 \alpha_{{\bf k}}^{\dag} u_{{\bf k}}^*(\tau) \Bigr)
\ee
where $[ \alpha_{{\bf k}}, \alpha_{{\bf k'}}^{\dag} ] = (2\pi)^3
\delta^3({\bf k} - {\bf k'})$ and $\alpha_{{\bf k}} | \textrm{in}
\rangle = 0$. Taking $| \textrm{in} \rangle$ to be the
Bunch-Davies (BD) vacuum, one finds the classic result
\be\label{generalmode}
 u_{\bf k}(\tau) = - \frac{\sqrt{\pi\tau}}{2a}
 \; H^{(1)}_\nu(-k\tau)
\ee
where $H^{(1)}_\nu(z)$ is a Hankel function, whose order is $\nu^2
= 9/4 - M^2/H^2$. For $|M^2| = |m^2 - 12 \,\xi H^2| \ll H^2$ this
becomes $\nu \simeq \frac32 - \epsilon$, with
\be
 \epsilon = \frac{M^2}{3H^2} = \frac{m^2}{3H^2} - 4 \, \xi \,,
\ee
satisfying $|\epsilon| \ll 1$.
In the case $M^2 = m^2 - 12 \xi H^2 = 0$, the modes simplify to
\be
 u_{\bf k}(\tau) = \frac{iH}{\sqrt{2k^3}} (1 + ik\tau)
 e^{-ik\tau} \,,
\ee
and the Green's functions are
\bea\label{BDmassless}
 G_\ssC^0(k,\tau_1,\tau_2) &=& - \frac i2 \,
 \Bigl( G^{-+} + G^{+-} \Bigr) \nn\\
  &=& \frac{H^2}{2k^3} \, \Bigl\{ (1 + k^2 \tau_1 \tau_2)
 \cos[ k(\tau_1 - \tau_2)] + k(\tau_1 - \tau_2) \sin [
 k (\tau_1 - \tau_2)] \Bigr\} \,, \nn\\
 &\simeq& \frac{H^2}{2 k^3} \Bigl\{ 1 + \cO[(k\tau)^2] \Bigr\}\,,
\eea
where the last line specializes to the long-wavelength,
super-Hubble limit,
\be
 -k \tau = \frac{k}{aH} \ll 1 \,.
\ee
%
%Here $\tau$ is conformal time, which is related to the de Sitter
%scale factor, $a$, and cosmological time, $t$, by $a = e^{H t} =
%-1/(H \tau)$.
The retarded correlator is similarly
\bea
 G_\ssR^0(k, \tau_1,\tau_2)
 &=& \theta(\tau_1 - \tau_2) \,
 \left( G^{-+} - G^{+-} \right) \nn\\
 &=& \theta(\tau_1 - \tau_2) \,
 \frac{H^2}{k^3} \, \Bigl\{ (1 + k^2 \tau_1 \tau_2)
 \sin[ k(\tau_1 - \tau_2)] - k(\tau_1 - \tau_2) \cos [
 k (\tau_1 - \tau_2)] \Bigr\} \nn\\
 &\simeq& \theta(\tau_1 - \tau_2) \,
 \frac{H^2}{3} (\tau_1^3 - \tau_2^3)
 \Bigl\{ 1 + \cO[(k\tau)^2] \Bigr\}  \,.
\eea
In these expressions the superscript `${0}$' distinguishes the
lowest-order, or `free', Green's function from the loop-corrected
one considered in later section.

\subsection{Divergences}

Some of the divergences arising in loop corrections are already
visible when the coincidence limit of the two point function is
evaluated in real space,
\be
 \Lambda(\tau) \equiv \Expect{\phi^2(x)} = - i G^{-+} (x,x) =
 G_\ssC^0(x,x) = \int \frac{\exd^3k}{(2\pi)^3}
 \, G_\ssC^0(k,\tau,\tau) +
 \rm{c.t.} \,,
\ee
where $\rm{c.t.}$ denotes the contribution of the counter-terms.
The retarded propagator does not contribute to this expression
since it vanishes at equal times. The isometries of de Sitter
space imply $\langle \phi^2(x) \rangle$ must be independent of $x$
if the expectation value is taken within any de Sitter-invariant
state (like the BD vacuum in particular) \cite{CandelasRaine}.

This expression diverges in both the UV and IR when evaluated for
a massless field in the BD vacuum. To regulate these we introduce
both an IR cutoff and a UV cutoff, with these cutoffs set (for
later convenience) at a fixed physical momentum scale:
$\Lambda_\IR < k/a < \Lambda_\UV$ (more about this choice below).
The regulated expression for $\Lambda(\tau)$ then becomes
\bea \label{gap}
 \Lambda(\tau) &=& \frac{1}{(2\pi)^2} \int_{a \,\Lambda_{\IR}}^{a\, \Lambda_\UV}
 \frac{\exd k}{k} \left\{ H^2 \left[ 1 + \left( \frac{k}{aH}
 \right)^2 \right]\right\} + \hbox{c.t.} \\
  &=& \frac{1}{(2\pi)^2} \left[ \int_{a\, \Lambda_{\IR}}^{a\, \Lambda_\UV}
 \frac{\exd k}{k} \left\{ H^2 \left[ 1 + \left( \frac{k}{aH}
 \right)^2 \right]\right\} - \int_{a \, \mu}^{a\,\Lambda_\UV}
 \frac{\exd k}{k} \left\{ H^2 \left[ 1 + \left( \frac{k}{aH}
 \right)^2 \right]\right\} \right] \,. \nn
\eea
As before, `c.t.' denotes the contribution of the counter-terms,
an explicit expression for which is given in the second line in a
particular renormalization scheme, associated with a
renormalization point $\mu$. We are left with the UV-finite result
\bea \label{gap2}
 \Lambda(\tau) &=& \frac{1}{(2\pi)^2} \left[ \int_{a\, \Lambda_{\IR}}^{a\, \mu}
 \frac{\exd k}{k} \left\{ H^2 \left[ 1 + \left( \frac{k}{aH}
 \right)^2 \right]\right\} \right] \\
  &\simeq& \frac{1}{(2\pi)^2}  \left[ H^2 \ln \left(
 \frac{\mu}{\Lambda_\IR} \right) + \frac12 \,
 (\mu^2 - \Lambda_\IR^2) \right]
 \,,\nn
\eea
Notice that $\Lambda(\tau)$ is $\tau$-independent, as would be
expected for a de Sitter invariant vacuum, when the divergence is
expressed in terms of a physical cutoff.

As always in quantum field theory, the appearance of an infrared
divergence tells us something important: since physical
observables must be infrared finite, we are not yet computing
something physical. If our calculation diverges we are not
describing the relevant long-distance physics with sufficient
accuracy, and so must have left out contributions whose presence
would cancel the IR divergence. But this means we are missing
contributions that must be just as important as those we compute.

Precisely what this missing IR physics is depends on the details
of the physical question being asked. For instance, if the IR
divergence arises in the scattering of charged particles, then the
missing physics could be related to the possibility of radiating
soft photons, meaning it was a mistake to choose the initial and
final states to have a definite number of photons ({\it \`a la}
Bloch-Nordsieck \cite{BN}). Alternatively should the divergence
arise when calculating an atomic energy level, it is resolved if
one uses atomic bound states to perform the calculation.
Similarly, for divergences arising at finite temperature
--- either at a critical point \cite{WK,RG} or in a hot plasma of
gauge particles \cite{HTL} --- the divergence indicates the
breakdown of the loop expansion. In this case loops are not simply
counted by powers of the coupling constant, and so the missing
physics that cures the divergence comes from higher-loop graphs
that are not suppressed relative to lower-loops by a small
parameter.

For the particular case of a logarithmic divergence, the
dependence on the missing IR physics is comparatively weak, and
more general things can be said about the final answer. Once the
appropriately modified IR part of the calculation is combined with
the result above, the logarithms of $\Lambda_\IR$ must cancel, as
in
\bea
 G_{\ssC}(k,L) &=& G_\ssC^\UV(\mu/\Lambda_\IR )
 + G_\ssC^\IR(\Lambda_\IR L) \nn\\
 &=& \Bigl[ A + B \ln \left(
 \frac{\mu}{\Lambda_\IR } \right) + \cdots\Bigr]
 + \left[C + B \ln \left(
 \Lambda_\IR L \right) + \cdots\right] \nn\\
 &=& (A + C) + B \ln \left(  \mu L \right) + \cdots \,,
\eea
to leave $\Lambda_\IR$ replaced by whatever the physical scale,
$L$, is that characterizes the infrared part of the real physical
observable. What is important about logarithmic divergences is
that this cancellation preserves the coefficient $B$, so that the
coefficient of the physical large logarithm, $\ln(\mu L)$, in the
full result can be computed purely by identifying the coefficient
of the IR-divergent logarithm within the UV theory. The same is
typically not also true for power-law divergences. As applied to
atomic energy levels this argument leads to the famous factor
\cite{Lambshift} of $\ln(m_e L) \simeq \ln (1/\alpha)$ in the Lamb
shift.

For the case of interest, $\langle \phi^2 (x) \rangle$ in de
Sitter space, whether the physical scale $L$ depends on time or
not depends on whether it breaks the de Sitter invariance. For
instance, $L$ is $\tau$-independent if the relevant IR physics is
a scalar mass. Alternatively, if it is an earlier pre-inflationary
phase that cuts off the long-distance de Sitter behavior, then $L$
would in general depend on $\tau$.

\subsection{Loop corrections}

\begin{figure} [ht]
\begin{center}
\includegraphics[width=0.9\textwidth,angle=0]{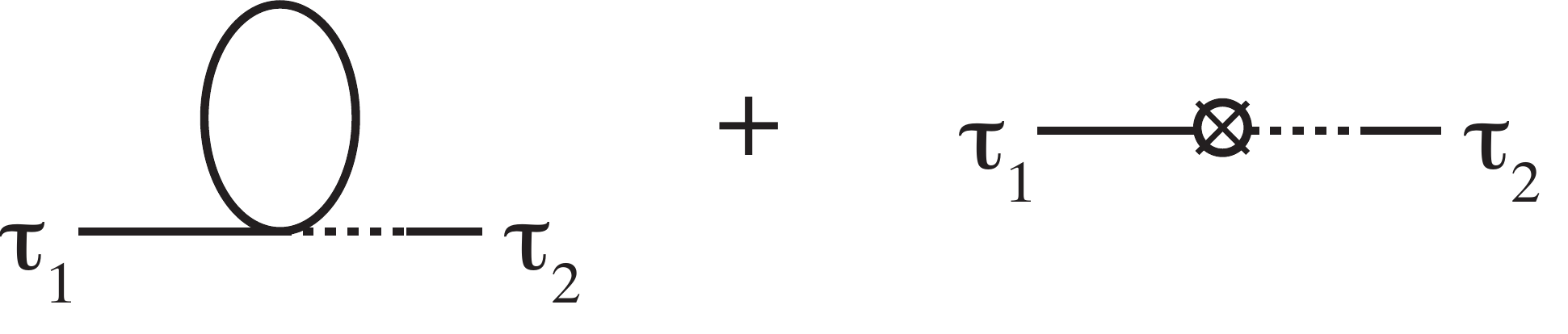}
\caption{The one-loop contribution to the $G_\ssC$ propagator. An
additional contribution comes from the related diagram with
$\tau_1$ and $\tau_2$ exchanged. Solid lines represent $G_\ssC$,
dashed-to-solid lines represent $G_\ssR$, and the crossed circle
denotes the counterterms.} \vspace{1cm}\label{fig:Quartic1}
\end{center}
\end{figure}

We now compute the loop corrections, basing our discussion on the
one-loop corrections to the propagator $G_\ssC$, focussing on the
contribution from super-Hubble scales. The relevant part of the
Lagrangian for this theory (after field doubling and changing to
the basis given by Eq.(\ref{Wmatrix})) is
\bea
 \mathcal{L}(\phi_C, \phi_\Delta) &=& \mathcal{L}(\phi^+)
 - \mathcal{L}(\phi^-) \nn\\
 &=&\sqrt{-g} \left[ %\hf \,
 g^{\mu\nu}\partial_\mu\phi_\ssC
 \partial^\nu \phi_\Delta + \frac{\lambda}{4!}
 \left(4\phi_\ssC^3\phi_\Delta + \phi_\ssC\phi_\Delta^3
 \right)
 + \rm{c.t.} \right]
\eea
Given the Feynman rules from this Lagrangian there are two
diagrams that contribute at next-to-leading order, one of which is
shown in Fig.~(\ref{fig:Quartic1}). The second diagram is obtained
from this one by interchanging $\tau_1 \leftrightarrow \tau_2$.

Evaluating Fig.~1 gives the following result for the 1-loop
correction \cite{Sloth,vMS, Petri}
\be\label{loop}
 G_\ssC^1(k,\tau_1,\tau_2) = -\frac{\lambda}{2}
 \int_{-\infty}^0 \exd\tau'
 a^4(\tau') \, G_\ssC^0(k,\tau_1,\tau') \, G_\ssR^0(k,\tau',\tau_2)
 \, \Lambda(\tau') + (\tau_1 \leftrightarrow \tau_2)
\ee
where the factor of $\frac12$ is from the combinatorics of the
contractions, and the quantity $\Lambda(\tau)$ is the momentum
integral given by Eq.~(\ref{gap2}). The contribution coming from
long-wavelength, super-Hubble modes is obtained by specializing to
that part of the integration region for which we can use the
asymptotic form for $-k \tau_{1,2} \ll 1$. Dropping $\cO(1)$ terms
relative to $\ln(-k\tau)$ gives
\bea
 G_\ssC^1(k,\tau_1,\tau_2) &\simeq& - \frac{\lambda}{2} \left(
 \frac{H}{2\pi} \right)^2  \frac{1}{6k^3} \int_{-1/k}^{\tau_1}
 \frac{\exd\tau'}{\tau'} \left[ \left( \frac{\tau_1}{\tau'}
 \right)^3 - 1 \right]
 \, \ln \left( \frac{\mu}{ \Lambda_{\IR}} \right)
 + (\tau_1 \leftrightarrow \tau_2)\nn\\
 &\simeq& \frac{\lambda}{2} \left(
 \frac{H}{2\pi} \right)^2  \frac{1}{6k^3}
 \, \ln \left( \frac{\mu}{ \Lambda_{\IR}} \right)
 \, [ \ln(- k \tau_1) + \ln(- k \tau_2) ]+\cdots\,.
\eea

Finally, evaluating at equal times, $\tau_1 = \tau_2 \equiv \tau$,
gives the following, IR-divergent, next-to-leading contribution to
$G_\ssC$:
\be \label{logform1}
 G_\ssC(k, \tau) = \frac{H^2}{2 k^3} \left[
 1 + \frac{\lambda}{3(2 \pi)^2} \ln \left( \frac{
 \mu}{\Lambda_\IR} \right) \ln \left( -k \tau \right)
  + \cdots \right] \,,
\ee
coming from super-Hubble modes. Here the ellipses denote terms whose contributions do
not diverge as either $\Lambda_\IR$ or $-k \tau$ vanish, and so
are subdominant to those explicitly displayed.

\subsection{Large logs and secular time dependence}

Eq.~\pref{logform1} displays two separate kinds of large
logarithms. The one of interest for the arguments of the next
section is the $\ln(-k\tau)$ term, since this displays the secular
time-dependence for which the dynamical renormalization group is
designed. This term is proportional to the number of $e$-folds
between horizon exit of the mode $k$ and the time $\tau$ where we
evaluate the result: $-k\tau = k/(aH) = a_k/a$. For calculations
of curvature perturbations resulting from inflation, this apparent
growth does not give rise to a physical divergence: the curvature
perturbation is conserved outside the horizon and in fact one may
follow the classical evolution of the modes shortly after horizon
crossing until horizon re-entry \cite{Seery:2007wf}.

The other large log is the IR divergence as $\Lambda_\IR \to 0$.
This has its origin both in the use of massless fields, and in the
eternity of de Sitter space that the above calculation implicitly
assumes holds in the past of $\tau$. As discussed earlier, this
divergence ultimately cancels some other dependence on
$\Lambda_\IR$ in a physical observable, and because the divergence
is logarithmic the cancellation of $\Lambda_\IR$ gives
\be \label{logform11}
 G_\ssC(k, \tau) = \frac{H^2}{2 k^3} \left[
 1 + \frac{\lambda}{3(2 \pi)^2} \ln \left(
 \mu  L \right) \ln \left( -k \tau \right)
  + \cdots \right] \,,
\ee
where $L$ is a physical scale associated with whatever
long-distance physics makes the result IR finite. The precise
nature of this scale is model-dependent, because it cannot be
identified purely within the UV part of the theory we use to this
point. We identify below what $L$ is in the specific case where it
is a mass that provides the convergence.

\section{RG Methods}

This section briefly recaps standard lowest-order RG arguments,
for their later use in our cosmological applications. Part of our
purpose is also to emphasize how the use of the RG equations
avoids the necessity of performing explicit summations over
infinite classes of graphs.

\subsection{The standard argument}

Standard one-loop calculations for dimensionless couplings give
the value of the renormalized couplings, $\alpha(\mu)$, for
different values of the renormalization point, $\mu$:
\be \label{logform}
 \alpha(\mu) = \alpha(\mu_0) + b \, \alpha^2(\mu_0) \,
 \ln \left( \frac{\mu}{\mu_0} \right) \,,
\ee
where $b$ is a calculable number. The domain of validity of this
calculation requires both $\alpha(\mu_0) \ll 1$ and $\alpha(\mu_0)
\ln(\mu/\mu_0) \ll 1$.

The RG proceeds \cite{RG} by differentiating the above result with
respect to $\mu$, giving
\be \label{difflogform}
 \mu \, \frac{\partial \alpha}{\partial \mu} = b \, \alpha^2 \,,
\ee
where, to within the accuracy given, $\alpha$ on both sides can be
taken to be $\alpha(\mu)$. This equation can be integrated to give
the solution
\be \label{difflogformsoln}
 \frac{1}{\alpha(\mu)} = \frac{1}{\alpha(\mu_0)} - b \,
 \ln \left( \frac{\mu}{\mu_0} \right) \,.
\ee
The point of this seemingly content-free exercise of
differentiating and then integrating the loop result is this:
while eq.~\pref{logform} relies on the product $\alpha
\ln(\mu/\mu_0)$ being small, eqs.~\pref{difflogform} and
\pref{difflogformsoln} only require $\alpha$ be small and so have
a larger domain of validity, applying in particular when $\alpha
\ln(\mu/\mu_0)$ is order unity. Integrating the RG equation
effectively resums the leading logs. Notice that this resummation
selectively takes terms to all orders in $\alpha \ln(\mu/ \mu_0)$,
but this does not rely on identifying which subclass of graphs is
dominant.

\subsection{The dynamical RG}

The dynamical RG \cite{DRG} uses a similar logic to identify the
long-time behavior of solutions when perturbative methods generate
perturbations that grow secularly with time. For instance, suppose
an approximation scheme based on expanding in a small parameter
$\varepsilon$ produces a result of the form
\bea
 y(t) &=& y_0(t)  + \varepsilon \, y_1(t) + \cO \left(\varepsilon^2 \right) \nn\\
 &=& y_0(c,t) + \varepsilon \, y_1(c, t) + \cO \left( \varepsilon^2 \right)\,,
\eea
where the second line emphasizes the dependence of the solution at
each order in $\varepsilon$ on the integration constant, $c$, of
$y_0(t)$. Of particular interest is when $y_1(t)$ is a secular,
growing function of time. When this is so, the expansion in powers
of $\varepsilon$ breaks down for time-scales, $t > T$, where
$\varepsilon y_1(T)/y_0(T)$ is order unity.

The dynamical RG shows how to infer the behavior of the solution
for times $t > T$. To do so, first introduce an arbitrary time
scale, $\vartheta$, with
\be
 y(t) = y_0(t) + \varepsilon \, \Bigl[ y_1(t) - y_1(\vartheta)
 + y_1(\vartheta) \Bigr] + \cO\left(\varepsilon^2 \right)  \,,
\ee
and absorb the last $y_1(\vartheta)$ term into the zeroeth-order
integration constants:
\be
 y_0[c(\vartheta), t] :=
 y_0(c,t) + \varepsilon \, y_1(\vartheta) \,,
\ee
and so
\be
 y(t) = y_0[c(\vartheta),t] + \varepsilon \, \Bigl[ y_1(t) - y_1(\vartheta)
 \Bigr] + \cO\left(\varepsilon^2 \right)  \,.
\ee

The dynamical RG argument proceeds from the recognition that
$y(t)$ is independent of $\vartheta$: $\exd y/\exd \vartheta = 0$.
Differentiating and dropping $\cO(\varepsilon^2)$ terms then gives
\be \label{Cdiff}
 \left( \frac{\partial y_0}{\partial c}
 \right) \, \frac{\exd c}{\exd \vartheta} - \varepsilon
 \, \frac{\partial y_1(c,\vartheta)}{\partial \vartheta} = 0 \,,
\ee
which can be regarded as a differential equation to be solved for
$c(\vartheta)$, whose solution is $c = \tilde c(\vartheta)$.

The argument is now the usual one: the solution $c = \tilde
c(\vartheta)$ has a broader domain of validity (typically simply
$\varepsilon \ll 1$) than does the initial definition of
$c(\vartheta)$ (which required $\varepsilon \, y_1(t) \ll 1$). The
final late-time behavior is then found by using the fact that
$y(t)$ does not depend on $\vartheta$ to make the convenient
choice $\vartheta = t$, and so
\bea
 y(t) &=& y_0[\tilde c(\vartheta),t]  + \varepsilon \, \Bigl[ y_1(t)
 - y_1(\vartheta) \Bigr] + \cO\left(\varepsilon^2 \right)\nn\\
 &=& y_0[\tilde c(t), t] +  \cO\left(\varepsilon^2 \right)  \,.
\eea
This result resums the leading large-time behavior beyond the
leading order perturbative result.
Particularly simple is the case where
\be
 y(t) = c\, \left[1 + \varepsilon \, f(t) + \cO\left(\varepsilon^2 \right)
 \right] \,,
\ee
where $y_0 = c$ is time-independent. In this case $c(\vartheta) =
c \, [1 + \varepsilon \, f(\vartheta)]$ and so the condition $\exd
y/\exd \vartheta = 0$ implies
\be
 \frac{\exd c}{\exd \vartheta} - \varepsilon \, c \, \frac{\exd
 f}{\exd \vartheta} = 0 \,,
\ee
which integrates to give $c = \tilde c(\vartheta)$, where $\tilde
c(\vartheta) = c \, e^{\varepsilon f(\vartheta)}$. This gives the
resummed large-time solution
\be
 y(t) = c \, e^{\varepsilon f(t)} \left[ 1 + \cO\left(\varepsilon^2 \right)
 \right] \,.
\ee
This is the particular case that arises in the one-loop scalar
calculation on de Sitter space.

\subsection{The dynamical RG for de Sitter fluctuations}

Earlier sections show the leading infrared-sensitive part of the
de Sitter space calculation of massless-scalar fluctuations
involves a secular time dependence,
\be \label{IRmasslessresult}
 G_\ssC(k, \tau) = \frac{H^2}{2 k^3} \left[
 1 + \frac{\lambda}{3(2 \pi)^2} \ln \left( \mu L
 \right) \ln \left( -k \tau \right) + \cdots
  \right]  \,,
\ee
with ellipses indicating terms that do not diverge when $L$ or $-k
\tau$ vanish. For this result, and assuming that the scale $L$ is
independent of $t$ --- as is the case if the IR physics is de
Sitter invariant (like a scalar mass term) --- the dynamical RG
resummation described above predicts the following long-time
behavior
\bea \label{DRGGC}
 G_\ssC(k, \tau) &=& \frac{H^2}{2 k^3} \exp \left[
  + \frac{\lambda}{3(2 \pi)^2} \ln \left( \mu L
 \right) \ln \left( -k \tau \right) \right] \Bigl( 1 + \cdots
 \Bigr) \nn\\
 &=& \frac{H^2}{2 k^3} \left( \frac{k}{aH} \right)^\delta
 \Bigl( 1 + \mathcal{O}(\delta^2)
 \Bigr)\,,
\eea
with
\be
 \delta = \frac{\lambda}{3(2 \pi)^2} \ln \left( \mu L
 \right)   \,.
\ee
This resummed result holds to all orders in $\delta \ln(- k\tau)$,
but neglects terms that are suppressed by additional powers of
$\delta$ without accompanying powers of $\ln(-k \tau)$. Its
validity therefore requires $\delta \ll 1$, although this
condition might also be relaxed through an RG resummation of the
large $\ln \left( \mu L \right)$ logarithm.

Notice that because $\mu L \gg 1$ and $\lambda > 0$ the power
$\delta$ is also positive, implying that the DRG-improved
correlator, eq.~\pref{DRGGC}, is less singular as $k \to 0$ than
is the zeroth-order result, $G^0_\ssC(k,\tau) \propto H^2/k^3$.
Also notice that although we do not need to know which subclass of
graphs is dominant in making this argument, the results of
ref.~\cite{Petri} show that these leading logs come from the
`chain' diagrams that repeatedly insert a scalar self-energy. The
validity of the DRG does not require that these entire diagrams
dominate all of the others, but only that they dominate in
their contribution to the leading logs.

\section{Comparisons to other results}

This section makes three points. First, the small-$k$ limit of the
above DRG expression for $G_\ssC(k,\tau)$ is argued to be similar
to the small-$k$ result for a massive scalar, showing that the
resummed late-time behavior softens the IR divergence in way very
similar to a mass. Second, the super-Hubble part of the loop
calculation is repeated using massive scalars, to show precisely
what the large logarithm $\ln(\mu L)$ means in an explicit example
for which the IR physics is known and calculable. Finally, we
extend our calculations to an $N$-scalar model for which a
dynamical mass is known to be generated within a controllable
approximation in the large-$N$ limit. We do so in order to show
more precisely how the DRG resummation reproduces the effects of
the dynamically resummed mass.

\subsection{Massive scalar field}

A massive scalar field in de Sitter space provides the simplest
example of IR-convergent, de Sitter invariant, physics which
nonetheless can have interesting contributions from super-Hubble
modes if the mass is less than the Hubble scale.
%$M^2_\ssR = m_\ssR - 12 \,\xi_\ssR H^2 \ll H^2$
Just as is true for flat space, a mass term improves the IR convergence by
making propagators less singular for small $k$. To see why this is
so for de Sitter space, recall that the massive scalar propagator
obtained from the general mode functions, eq.(\ref{generalmode}),
becomes, in the limit $-k\tau = k/(aH) \ll 1$
\bea\label{massive}
 G_\ssC^0(k,\tau_1,\tau_2) & \simeq &
 \frac{H^2}{2k^3} (k^2\tau_1\tau_2)^\epsilon\\
 G_\ssR^0(k,\tau_1,\tau_2)  & \simeq & \theta(\tau_1-\tau_2)
 \frac{H^2}{3}(\tau_1^{3-\epsilon} \tau_2^\epsilon-
 \tau_1^\epsilon\tau_2^{3-\epsilon}) \,.
\eea
where $\epsilon = {M^2}/{3H^2}$.  As advertised, these expressions
cure the IR divergences encountered previously because they are
less singular than the massless case as $k \to 0$. But for $M^2
\ll H^2$ the growth of the difference between the massive and
massless expressions for $G_\ssC(k,\tau)$ is much slower than in
flat space, with significant deviations only arising once $(-k
\tau)^{2\epsilon}$ deviates from unity. This occurs for $k < k_*$
with
\be
 -k_* \tau = \frac{k_*}{aH} \simeq e^{-1/2\epsilon} = e^{-3H^2/2M^2}
 \,,
\ee
and so $k_*/a \simeq H \, e^{-3H^2/2M^2}$ can be much smaller than
$M$.

What is important for the present purposes is that the small-$k$
form of the massive propagator has the same kind of $k$-dependence
as does the DRG-resummed result for the massless field: both
behave as $G_\ssC \propto k^{\alpha - 3}$, for $\alpha
> 0$. For massive particles we have $\alpha = 2 \epsilon =
2 M^2/ 3 H^2$ while DRG resummation gives $\alpha = \lambda
\ln(\mu L)/12 \pi^2$. Equating these powers shows that the
DRG-resummed long-wavelength correlator behaves {\em as if}\ it
describes a dynamically generated mass of order
\be
 M^2_{\rm eff} \simeq \frac{\lambda H^2}{8 \pi^2} \, \ln (\mu L) \,.
\ee
A reasonable guess for the size to take for $L$ in this expression
is
\be \label{Lmguess}
 L \simeq \frac{a}{k_*} \simeq \frac{1}{H} \; e^{1/2\epsilon}
\,, \ee
(a more precise choice for $L$ is described shortly), in which
case we have
\be
 M^2_{\rm eff} \simeq \frac{\lambda H^2}{8 \pi^2} \, \ln \left(
 \frac{\mu}{H} \right) + \frac{3 \lambda H^4}{16 \pi^2 M^2} \,.
\ee

\subsection{Massive loop calculation}

It is instructive to repeat the IR-sensitive, super-Hubble part of
the loop calculation using this same massive scalar (still with a
quartic self-interaction), since this provides a simple and
specific example of IR-safe physics that can cure the IR
divergences of the massless case. Among other things, such a
calculation allows us to identify more precisely what the large
logarithm $\ln(\mu L)$ means when it is a
small mass that is responsible for IR convergence.

So far as the purpose of identifying the IR behavior is
concerned, what is important is following the contribution of the
super-Hubble modes in the result. For the loop factor,
$\Lambda(\tau)$, the contribution of these modes is
\bea \label{massiveLambda1}
 \Lambda(\tau) &\simeq& \frac{ H^2 }{(2\pi)^2}
 \int_{a\, \Lambda_{\IR}}^{a \,\Lambda_\UV}
 \frac{\exd p}{p} \left(- p \,\tau
 \right)^{2\epsilon}  + \hbox{c.t.}
  \simeq \frac{H^2}{(2\pi)^2}
  \int_{a\, \Lambda_{\IR}}^{a\, \mu}
 \frac{\exd p}{p} \; \left( -p \,\tau
 \right)^{2\epsilon}  \nn\\
 &\simeq& \frac{1}{2\epsilon} \left(
 \frac{H}{2\pi} \right)^2  \left[ \left( \frac{\mu}{H}
 \right)^{2\epsilon} - \left( \frac{\Lambda_\IR}{H}
 \right)^{2\epsilon} \right] \,,
\eea
which is again $\tau$-independent because of the choice that
$\Lambda_\IR$ and $\mu$ are physical scales. Notice that in the
limit $\epsilon \to 0$ eq.~\pref{massiveLambda1} approaches the
super-Hubble part of the massless case, eq.~\pref{gap2},
\be
 \lim_{\epsilon \to 0} \Lambda = \left( \frac{H}{2\pi} \right)^2
 \ln \left( \frac{\mu}{\Lambda_\IR} \right) \,,
\ee
but instead converges in the IR if $\epsilon$ is held fixed:
\be \label{massiveLambda}
  \lim_{\Lambda_\IR \to 0} \Lambda = \frac{1}{2\epsilon} \left(
 \frac{H}{2\pi} \right)^2  \left( \frac{\mu}{H}
 \right)^{2\epsilon} \,.
\ee
Here $\mu$ is, as before, a renormalization scale associated with
the UV counterterms.
Using eq.~\pref{massiveLambda} to evaluate the 1-loop result,
Fig.~1, then gives
\bea
 G_\ssC^1(k,\tau_1,\tau_2) &=& -\frac{\lambda}{2} \int \exd\tau'
 a^4(\tau') \, G_\ssC^0(k,\tau_1,\tau') \, G_\ssR^0(k,\tau',\tau_2)
 \, \Lambda(\tau') + (\tau_1 \leftrightarrow \tau_2) \nn\\
  &=& -\frac{\lambda}{2} \left[ \frac{1}{2\epsilon}
  \left( \frac{H}{2\pi} \right)^2 \left( \frac{\mu}{H}
 \right)^{2\epsilon} \right] \frac{1}{6k^3}
  \int_{-1/k}^{\tau_1} \frac{\exd\tau'}{\tau'}
  \; \left[
   \left( \frac{\tau_1}{\tau'} \right)^{3-\epsilon}  -
   \left( \frac{\tau_1}{\tau'} \right)^{\epsilon} \right]
    (k^2 \tau' \tau_2)^{\epsilon} \nn\\
  &&\qquad\qquad\qquad\qquad\qquad\qquad\qquad\qquad\qquad\qquad
  + (\tau_1 \leftrightarrow \tau_2)\nn\\
  &\simeq& \frac{\lambda}{4\epsilon}
  \left( \frac{H}{2\pi} \right)^2  \frac{1}{6k^3}
  \left( \frac{k^2 \mu^2 \tau_1 \tau_2}{H^2} \right)^\epsilon
  \left[ \ln \left( -k \tau_1 \right) +\ln \left( -k \tau_2 \right)
   + \cdots \right]
  \,,
\eea
where the unwritten terms are finite as $k\tau \to 0$. We find in
this way the loop-corrected result
\be\label{massiveloops}
 G_\ssC(k,\tau) \simeq \frac{H^2}{2k^3} (-k\tau)^{2\epsilon}
 \left[ 1 + \frac{\lambda}{6 (2\pi)^2 \epsilon} \, \left(
 \frac{\mu}{H} \right)^{2\epsilon}
 \ln(-k\tau) + \cdots \right] \,.
\ee
Notice that this is similar to expression \pref{IRmasslessresult}
in the massless case, with the replacement
\be
 \ln ( \mu L) \to  \frac{1}{2\epsilon} \left( \frac{\mu}{H}
 \right)^{2 \epsilon} = \frac{3 H^2}{2 M^2} \, \left(
 \frac{\mu}{H} \right)^{3M^2/3H^2} \,.
\ee
For small $\epsilon$ this reduces to
\be
 \ln (\mu L) \simeq \frac{1}{2 \epsilon} + \ln\left(
 \frac{\mu}{H} \right) + \cO(\epsilon^2) \,,
\ee
which agrees with the simpler estimate, eq.~\pref{Lmguess}, of the
previous section. In this case the large logarithm $\ln(\mu L)$ is
seen to correspond in the full result to the enhancement of the
loop contribution by the factor $1/\epsilon$.

\subsubsection*{DRG resummation}

Applying the DRG to resum the leading logs gives in this case the
late-time result
\be
 G_\ssC(k,\tau) \simeq \frac{H^2}{2k^3} (-k\tau)^{2\epsilon + \delta_m}
\ee
where
\be \label{deltam}
 \delta_m = \frac{\lambda}{6(2\pi)^2 \epsilon} \left(
 \frac{\mu}{H} \right)^{2\epsilon} \,,
\ee
confirming that the resummed leading logs modify $G_\ssC(k,\tau)$
in the same way as would a mass. Notice that although this resums
terms to all order in $\delta_m \ln (-k \tau)$, it requires
$\delta_m \ll 1$, and so breaks down once $\lambda/\epsilon$
becomes too small.
When both masses and couplings are present the mass effect
dominates the coupling effect so long as $\delta_m < 2 \epsilon$,
or
\be
 \frac{\lambda}{(4\pi)^2} < 3 \epsilon^2 \left( \frac{H}{\mu}
 \right)^{2\epsilon} = \frac{M^4}{3 H^4}
 \left( \frac{H}{\mu} \right)^{2M^2/3H^2} \,.
\ee
For $\lambda$ larger than this it is the coupling that dominates
in the IR, and it contributes in the same way as would a mass of
order
\be \label{Meff}
 M^2_{\rm eff} = \frac{3H^2}{2} \; \delta_m
 = \frac{\lambda H^2}{(4\pi)^2 \epsilon} \left(
 \frac{\mu}{H} \right)^{2\epsilon}
 \simeq \frac{3\lambda H^4}{(4\pi)^2 M^2}  \,,
\ee
where the last approximation neglects corrections that are of
order $\epsilon$, since the assumption that the coupling effect
dominates requires $\epsilon^2$ to be systematically small
relative to $\lambda/(4\pi)^2$.

Notice that eq.~\pref{Meff} precisely agrees with what would be
expected in a mean-field approximation, for which\footnote{For the
numerical factor, compare with the large-$N$ expression in the
next section.} $M^2_{mf} \simeq \frac12 \, \lambda \langle \phi^2
\rangle$, given the nonzero expectation
\be
 \langle \phi^2 \rangle \simeq \frac{ 3 H^4}{8 \pi^2 M^2} \,,
\ee
that a massive scalar field prepared in the Bunch-Davies vacuum
acquires in de Sitter space \cite{Dowker:1975xj,Linde:1984ir}. To
this extent the DRG-improved long-time limit agrees with what
would be expected from a stochastic approach \cite{Starobinsky} to
super-Hubble scalar fluctuations. We explore this in more detail
using a large-$N$ limit \cite{Starobinsky, RS, Petri}
in the next section.

\subsection{Dynamical mass at large $N$}

Large-$N$ expansions provide a useful laboratory for understanding
dynamical resummations within a controlled approximation, and so
we therefore pause to examine them briefly here. In this section
we apply the dynamical RG to the large-$N$ generalization of the
above calculation in order to show how it compares with the
long-time behavior predicted by the DRG.

Consider to this end the large-$N$ generalization of our model,
\bea
 S &=& - \int \exd^4x \; \sqrt{-g} \; \left[ \frac12 \,
 g^{\mu\nu} \, \partial_\mu \Phi^\ssT \, \partial_\nu \Phi + V_0 +
 \frac12 \Bigl( m^2 + \xi R \Bigr)\, (\Phi^\ssT \Phi) +
 \frac{\lambda}{4!} \, (\Phi^\ssT \Phi)^2 \right] \nn\\
 &=& - \int \exd^4x \; \sqrt{-g} \; \left[ \frac12 \,
 g^{\mu\nu} \, \partial_\mu \Phi^\ssT \, \partial_\nu \Phi + V_0
 +  \frac{M^2}{2} \, (\Phi^\ssT \Phi)
  - \frac{\lambda}{4!} \left( s^2 - 2 s \, \Phi^\ssT \Phi
 \right) \right]\,, \nn\\
\eea
where $M^2 = m^2 - 12 \, \xi H^2$ in de Sitter space, $\Phi$
denotes an $N$-component column vector of real scalar fields, and
the first line follows from the second if $s$ is integrated out
using the exact result
\be
 s = \Phi^\ssT \Phi \,.
\ee

%\subsubsection*{Dynamical mass generation}

The scalar field equation for $\Phi$ in this model is
\be
 - \Box \Phi + \left( M^2 + \frac{\lambda s}{6} \right)\, \Phi = 0  \,.
\ee
In the limit of large $N$ we take $\lambda$ small and $\langle s
\rangle$ large, so that $g = N\lambda$ and $\sigma = \langle s
\rangle /N$ are fixed. In this large-$N$ limit the field equation
behaves sensibly, taking the mean-field form \cite{largeN}
\be
 - \Box \Phi + \left[ M^2 + \frac{g \sigma}{6}
 \right]\, \Phi = 0  \,,
\ee
which uses the large-$N$ expression $\lambda s \, \Phi = g \sigma
\, \Phi \left( 1 + 2/N \right) + \cdots$.  This shows that the
scalar self-interaction contributes to the mass for $\Phi$ in the
large-$N$ limit when terms of order $1/N$ are dropped
\be \label{MlargeN}
 M^2_\ssN = M^2 + \frac{g \sigma}{6} = m^2 - 12 \, \xi H^2
 + \frac{g \sigma}{6} \,,
\ee
provided only that $\sigma$ is nonzero.

The static value for $\sigma$ can also be computed for large $N$
in de Sitter space, with a standard result given by the
contribution of $N$ free scalar fields
\cite{Dowker:1975xj,Linde:1984ir}:
\be
 \sigma = \frac{1}{N} \, \langle \Phi^\ssT \Phi \rangle
 = \frac{3 H^4}{8 \pi^2 M^2} \,,
\ee
and so
\be \label{MlargeN1}
 M^2_{\ssN} = M^2 + \frac{g \sigma}{6}
 = M^2 + \frac{g H^4}{(4 \pi)^2 M^2} \,.
\ee
Using this mean-field mass in the $\Phi_\ssC$ correlator then
gives the large-$N$ expectation for the form of the small-$k$
limit of the de Sitter correlator
\be\label{massiveloops3}
 G_\ssC(k,\tau) \simeq \frac{H^2}{2k^3} (-k\tau)^{2\epsilon_{\ssN}}
 \,,
\ee
where
\be \label{epsN}
 2 \epsilon_{\ssN} = \frac{2 M^2_{\ssN}}{3H^2}
 = \frac{2 M^2}{3H^2} + \frac{gH^2}{6(2 \pi)^2M^2} \,.
\ee

\subsubsection*{DRG resummation}

For comparison we instead compute $G_\ssC(k, \tau)$ perturbatively
in $g$, and repeat the loop calculation of the previous section in
the large-$N$ limit. There are only two changes required. One
change comes from evaluating the combinatorial factors arising
from performing the contractions of the various fields appearing
in Fig.~1, which converts a factor of 3 (from counting the number
of ways to contract fields) in the previous loop calculation into
$(2 + N)$. The second change is the replacement of the coupling,
$\lambda \to g/N$, and so the generalization to arbitrary $N$ of
the previously obtained loop result is obtained by making the
simple replacement
\be
 3 \lambda \to g \left( 1 + \frac{2}{N} \right) \,.
\ee
This leads to the following large-$N$ generalization of
eq.~\pref{massiveloops},
\be\label{massiveloopsN}
 G_\ssC(k,\tau) \simeq \frac{H^2}{2k^3} (-k\tau)^{2\epsilon}
 \left[ 1 + \frac{g}{18 (2\pi)^2 \epsilon} \, \left(
 \frac{\mu}{H} \right)^{2\epsilon}
 \ln(-k\tau) + \cdots \right] \,.
\ee
Using the DRG to resum the large-$\tau$ behavior gives, as before
\be\label{DRGmN}
 G_\ssC(k,\tau) \simeq \frac{H^2}{2k^3}
 (-k\tau)^{2\epsilon + \delta_m}
 \,,
\ee
where
\be \label{deltamN}
 \delta_m = \frac{g}{18(2\pi)^2 \,\epsilon} \left(
 \frac{\mu}{H} \right)^{2\epsilon}
 \simeq \frac{g H^2}{6(2\pi)^2 M^2} \,,
\ee
and the second, approximate, expression uses $\epsilon \ll 1$ to
drop the factor $(\mu/H)^{2\epsilon}$ (compare with
eq.~\pref{deltam}) . For small $k$ the total power of $k$
therefore is
\be
 2\epsilon + \delta_m = \frac{2 M^2}{3H^2}
 + \frac{g H^2}{6(2\pi)^2
 M^2}\,,
\ee
in exact agreement with $\epsilon_\ssN$ obtained using the
large-$N$, mean-field result, eq.~\pref{epsN}.

\subsubsection*{Self-consistent masses}

The large-$N$ limit also allows a more careful exploration of the
case of very small $M^2$. Because the dynamical mass, $M_\ssN^2$
of eq.~\pref{MlargeN1}, is bounded from below, the physical mass
does not disappear even in the limit $M^2 \to 0$. Instead it
reaches its minimum value when $\partial M_\ssN^2 / \partial M^2 =
0$, or
\be \label{mdyn}
 M^2_{\rm min} =
 \frac{1}{4\pi} \sqrt{g} \; H^2 \,.
\ee
Because $g/\epsilon \simeq \sqrt g$ for this value of $M^2$ it
lies within the domain of validity of our approximate expressions,
which require $g/\epsilon \ll 1$. The subsequent growth of
$M^2_\ssN$ for smaller $M^2$ than this is suspect because
$g/\epsilon$ becomes larger than unity. This means that
eq.~\pref{mdyn} corresponds to the smallest mass that it is
possible to get within the domain of validity of our
approximations. (The corresponding result for $N=1$ would be
$M^2_{\rm min} = \sqrt{3\lambda} \; H^2/4\pi$. This mass is one
that has been obtained by solving a `gap' equation, in
ref.~\cite{Petri}, and agrees with the much earlier result of
Starobinsky and Yokoyama \cite{Starobinsky}, after rescaling to
match their definition of $\lambda$.)

\section{Scalar field with a cubic interaction}

In this section we consider a slightly different case: a scalar
field with a purely cubic interaction
\be
 V(\phi) = \frac{h}{3!} \, \phi^3 \,,
\ee
with no mass term and no quartic interaction.
%
%We do so because the DRG can be applied in this
%case much as before, even though the theory is not expected to
%generate a dynamical mass in the same way as occurred above.
%
We find in this case that the DRG-improved super-Hubble
contributions do {\em not} have a momentum dependence appropriate
to a dynamically generated mass.

%This example serves as a check that treating the IR scale $L$ as constant in time gives a consistent physical conclusion for both quartic and cubic interactions.

%\vspace{0.4in}
\begin{figure}[h]
\begin{center}
$\begin{array}{ccc}
\includegraphics[width=0.3\textwidth,angle=0]{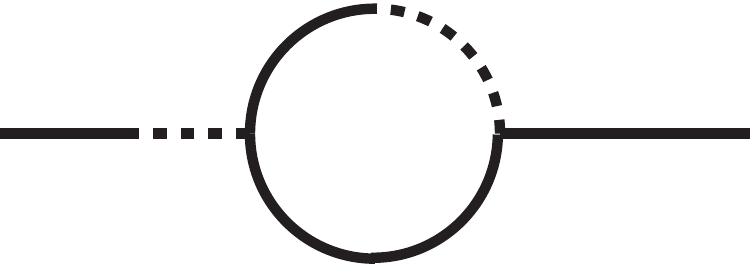} &
\includegraphics[width=0.3\textwidth,angle=0]{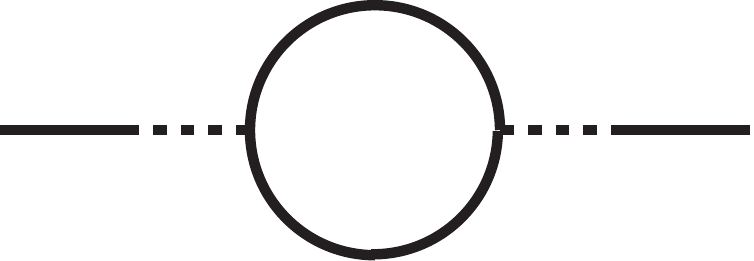} &
\includegraphics[width=0.3\textwidth,angle=0]{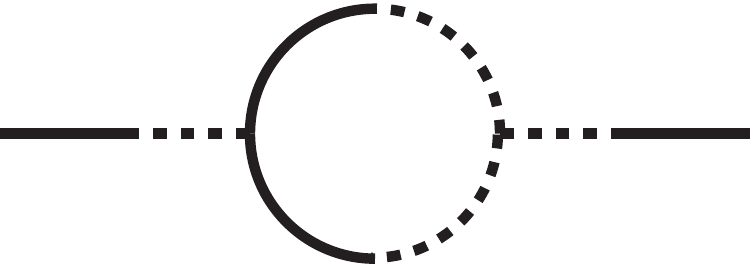}\\
\mbox{(a)}&\mbox{(b)}&\mbox{(c)}
\end{array}$
\caption{The three contributions to $G_c(k,\tau_1,\tau_2)$ at one
loop. Diagram (a) is divergent in the UV and IR, (b) is only IR
divergent, and (c) diverges only in the UV. }
\label{CubicDiagrams}
\end{center}
\end{figure}

The one-loop graphs (other than the counterterms) contributing to
$G_\ssC(k, \tau_1, \tau_2)$ in this theory are shown in Figure 2.
We evaluate these graphs\footnote{The complete Feynman rules for
this theory in the real time formalism are given in \cite{vMS}.
The left vertex of Figure 2(a) contributes
$-iha^4(\tau_1)\delta(\tau_1-\tau^{\prime})\delta(\tau_1-\tau^{\prime\prime})$,
where $\tau_1$ is the time associated with the incoming line and
$\tau^{\prime}, \tau^{\prime\prime}$ are associated with the lines
to the right of the vertex. This is the only new rule needed to
evaluate the IR divergent contributions in this example.} using
only the long-wavelength, super-Hubble limit of the propagators,
as before. All graphs contribute a factor of $[\ln(-k\tau)]^2$
thanks to the two interaction vertices, but each has a different
structure of divergences in the momentum integral. The first
diagram gives a contribution that grows in time and is
proportional to $\Lambda = (H/2\pi)^2 \ln(\mu/\Lambda_\IR)$, as in
the $\phi^4$ case. The second diagram converges in the UV and the
momentum integral is dominated by loop momenta $p\gg k$. The last
graph is UV divergent but not IR singular, and so does not
contribute to the terms of present interest.

Evaluating the contribution of the first and second graphs gives the
equal-time result
\bea
 G_\ssC(k,\tau, \tau) &=& G_\ssC^{0}(k,\tau,\tau)
 \left\{ 1+\frac{h^2}{9H^4}
 \left[ \ln^2(-k\tau)+\frac{4}{3}\ln(-k\tau)\right.\right.\\\nonumber
 &&\left.\left.\qquad\qquad\qquad
 +\frac{2}{3}(-k\tau)^3
 \ln(-k\tau)+\frac{2}{9}\right]\left[4\Lambda+\frac{H^2}{(2\pi)^2}\ln\left(\frac{k}{a\Lambda_{IR}}\right)\right]\right\}\\\nonumber
 &=&G_\ssC^{0}(k,\tau,\tau)\left\{ 1+\frac{h^2}{9(2\pi)^2H^2}\ln^3(-k\tau)+\frac{4h^2\Lambda}{9H^4}\ln^2(-k\tau)+\dots\right\}
 \,,
\eea
where $\Lambda(\tau) = (H/2\pi)^2 \ln(\mu/\Lambda_\IR)$ is as
defined in the previous section, and the second line uses $k/(a \,
\Lambda_{IR}) = (-k\tau)(H/\Lambda_{IR})$. A massive scalar field
with a cubic interaction is treated in detail in \cite{vMS}, who
also do a careful analysis of the UV pieces. However, the IR part
of the calculation above is sufficient to contrast the result of
the DRG for this case with the quartic interaction result.

Applying the DRG method to this expression exponentiates the
quantity in parenthesis,
\bea
 G_\ssC(k,\tau, \tau) &=& G_\ssC^{0}(k,\tau,\tau)
 \exp \left\{\frac{h^2}{9H^2}
 \left[\frac{1}{(2\pi)^2}\ln^3(-k\tau)+
 \frac{4\Lambda}{H^2}\ln^2(-k\tau)+\dots\right]\right\}
 \,, \nn\\
\eea
which clearly does not share the small-$k$ dependence appropriate
to a constant mass. This agrees with naive expectations, since
even if a mean-field treatment were valid it would predict
$M^2_{\rm dyn} \propto h \langle \phi \rangle = 0$. But unlike the
quartic case, for the purely cubic theory the semiclassical
expansion about the solution $ \langle \phi \rangle = 0$ is
unstable, making the study of its fluctuations problematic.

\section{Conclusions}

Loop calculations involving massless scalars in de Sitter space
have long been known to be plagued by strong infrared effects.
These arise due to the indefinite growth of fluctuations on
super-Hubble scales. These long-wavelength effects complicate the
precision calculation of inflationary observables, by potentially
putting the late-time behavior beyond the reach of perturbative
calculations.

We argue here that the DRG is a natural tool to use for this kind
of problem, where secular time dependence limits the domain of
validity of a perturbation expansion. It provides a simple way to
resum these terms, thus allowing the late-time behavior to be
extracted from perturbative calculations.

We apply the DRG to several simple examples, such as a single
quartically self-coupled field as well as a large $N$ quartically
coupled model. In these situations the resummed IR behavior is
less singular than is the leading-order result, in a way that
mimics the presence of a nonzero mass. The same analysis for a
cubic scalar theory demonstrates that the cubic interaction apparently
{\em does not} self-regulate by generating a mass.

Fortified by these checks, we believe that DRG is likely to prove
to be very useful for more detailed studies of other puzzling
aspects of late-time, long wavelength de Sitter fluctuations.

Another direction to take this work might be to consider
situations where there the important long-distance physics breaks
de Sitter invariance. This could happen because inflation is not
eternal, instead beginning at the end of a pre-inflationary phase
\cite{preHIosc}, or perhaps if dS space should prove to be only
metastable, with the BD vacuum ultimately decaying on large scales
due to the inhomogeneous structure, perhaps coming from the
formation of bubbles of some new phase. Whatever this source of dS
symmetry breaking is, the two point function becomes a function of
time and one may wonder whether a DRG analysis can still be
useful.

An interesting application of this type would be to reproduce past
discussions of such secular behaviour, such as that of
ref.~\cite{Woodard2} which discusses large infrared logarithms in
a time-dependent situation, and argues, following a proposal of
Starobinsky and Yokoyama \cite{Starobinsky}, that these may be
resummed \cite{Tsamis} using ideas from stochastic inflation.
Investigations of this point, and how it relates to the DRG
techniques discussed here, are in progress.  Previous arguments about applying
RG techniques for the case where the dS symmetry is broken can be found in
\cite{WoodardRG}.

%Note that we have found, in agreement with \cite{vMS}, that the
%classical equation of motion do capture the leading infrared
%behavior. In \cite{vMS}, they argue that solving the equation of
%motion perturbatively captures all the secularly growing terms and
%if one was able to solve the whole non-linear equation, it would
%amount to a resummation of all these terms. The use of DRG
%techniques is more general and agrees with a complete solution of
%the equation of motion when it is available.

\section*{Acknowledgements}

We would like to thank Jason Kumar, Tomislav Prokopec, Arvind
Rajaraman, David Seery and Andrew Tolley for many useful
conversations. We wish to thank the Aspen Center for Physics, for
providing the spectacular environment where this work was started,
and RH thanks the Perimeter Institute for hospitality as work
progressed. He also thanks the DOE for support through Grant  No.
DE-FG03-91-ER40682. This research has been supported in part by
funds from the Natural Sciences and Engineering Research Council
(NSERC) of Canada, and Perimeter Institute. Research at the
Perimeter Institute is supported in part by the Government of
Canada through NSERC and by the Province of Ontario through the
Ministry of Research and Information (MRI).

\appendix

\section{Renormalization}

When renormalizing the one-loop UV divergences it is important to
keep all of the interactions whose counter-terms can be required
to cancel divergences at one loop. For an $O(N)$-invariant system
of $N$ scalar fields the required terms are:
\bea \label{ctapp}
 S &=& - \int \exd^4x \; \sqrt{-g} \; \left[ \frac12 \,
 g^{\mu\nu} \, \partial_\mu \Phi^\ssT
 \, \partial_\nu \Phi + \frac12 (m^2 + \xi \, R)
 (\Phi^\ssT \Phi) + \frac{\lambda}{4!}
 \, (\Phi^\ssT \Phi)^2 \right. \\
 && \qquad\qquad\qquad\qquad \left. \phantom{\frac12}
 + V_0 + A \, R + B \, R_{\mu\nu\lambda\rho}
 R^{\mu\nu\lambda\rho} + C \, R_{\mu\nu} R^{\mu\nu}
 + D\, R^2 + E \, \Box R
  \right] \,. \nn
\eea
The coefficients of these terms contain divergent contributions,
whose presence cancels all UV divergences that arise at one loop.
The counterterm lagrangian consists of the same terms with $m^2
\to \delta m^2$, $\xi \to \delta \xi$ and so on.

Working in dimensional regularization, a standard calculation
\cite{GdW} gives the divergent part of the one-loop effective
action, $\Sigma_\infty$, generated by an arbitrary collection of
fields. Assuming these fields have an action of generic form
\be
 S = -\frac12 \int \exd^4 x \sqrt{-g} \;
 \Phi^\ssT (-\Box + X ) \Phi \,,
\ee
one finds\footnote{This can be written in an alternative form also
involving the lower-order Gilkey coefficients $a_0$ and $a_1$ by
grouping terms involving different powers of $m^2$.}
\be
 \Sigma_\infty = \pm\frac12 \, \left(
 \frac{1}{4\pi} \right)^{n/2}
 \Gamma(2 - n/2) \, \mu^{n-4}
 \int \exd^n x \sqrt{-g}
 \; {\rm tr} \, a_2 \,,
\ee
where the upper (lower) sign applies for boson (fermions), $n \to
4$ denotes the dimension of spacetime, $\mu$ is the arbitrary mass
scale that arises in dimensional regularization, and $a_2$ is the
second Gilkey/de Witt coefficient,
\bea \label{a2form}
 a_2 &=& \frac{1}{360} \Bigl( 2 R_{\mu\nu\lambda\rho}
 R^{\mu\nu\lambda\rho} - 2 R_{\mu\nu} R^{\mu\nu}
 + 5 R^2 - 12 \Box R \Bigr) \nn\\
 && \qquad\qquad + \frac16 \, R \, X + \frac12 \, X^2
 - \frac16 \, \Box X + \frac{1}{12} Y_{\mu\nu} Y^{\mu\nu} \,.
\eea
Here $[D_\mu, D_\nu] \Phi = Y_{\mu\nu} \Phi$ defines $Y_{\mu\nu}$,
where $D_\mu$ is the covariant derivative appearing within $\Box
\Phi = g^{\mu\nu} D_\mu D_\nu \Phi$.

Since gauge-neutral scalars are the case of present interest, we
have $Y_{\mu\nu} = 0$. Further specializing to the potential $V =
\frac12 (m^2 + \xi \, R) (\Phi^\ssT \Phi) + \frac{1}{4!} \,
\lambda (\Phi^\ssT \Phi)^2$ then gives
\be
 X = m^2 + \xi \, R + \frac{\lambda}{6} \, (\Phi^\ssT \Phi)
 + \frac{\lambda}{3} \, \Phi \, \Phi^\ssT \,,
\ee
and so
\be
 {\rm tr} \, X = N(m^2 + \xi \, R) +
 \frac{\lambda}{6} (\Phi^\ssT \Phi)  \left(
 N + 2 \right) \,,
\ee
and
\be
 {\rm tr} \, X^2 = N ( m^2 + \xi R)^2
 + \frac{\lambda}{3} \,  (m^2 + \xi R) (\Phi^\ssT \Phi)
 \left( N + 2 \right)
 + \frac{\lambda^2}{36} (\Phi^\ssT \Phi)^2
 \left( N + 8 \right) \,.
\ee
Using these in eq.~\pref{a2form} gives the result
\bea \label{a2}
 {\rm tr} \, a_2 &=& \frac{N}{360} \Bigl( 2 R_{\mu\nu\lambda\rho}
 R^{\mu\nu\lambda\rho} - 2 R_{\mu\nu} R^{\mu\nu}
 + 5 R^2 - 12 \Box R \Bigr) \nn\\
 && \qquad\qquad + \frac16 \, R \; {\rm tr} \, X
 + \frac12 \; {\rm tr} \, X^2
 - \frac16 \; \Box \, {\rm tr} \, X  \nn\\
 &=& \frac{N}{180} \Bigl( R_{\mu\nu\lambda\rho}
 R^{\mu\nu\lambda\rho} - R_{\mu\nu} R^{\mu\nu} \Bigr)
 + \frac{N}{2} \left( \xi + \frac16 \right)^2
 R^2 - \frac{N}{6} \left( \xi + \frac15 \right) \Box R \nn\\
 && \qquad\qquad + \frac{N m^4}{2} + N \left( \xi + \frac16
 \right) m^2 R + \frac\lambda6 (N + 2) \, m^2 ( \Phi^\ssT \Phi)
 \nn\\ && \qquad\qquad
 + \frac\lambda6 (N+2) \left( \xi + \frac16 \right)
 R \, ( \Phi^\ssT \Phi) + \frac{\lambda^2}{72} (N+8)
 ( \Phi^\ssT \Phi)^2 \nn\\
 && \qquad\qquad - \frac{\lambda}{36} \, (N+2)
 \, \Box ( \Phi^\ssT \Phi ) \,.
\eea

Using the above results to compute $\Sigma_\infty$, shows that the
one-loop divergences are given by poles as $n \to 4$ of the form
\be
 \Sigma_\infty = -\left(
 \frac{1}{4\pi} \right)^{n/2} \left( \frac{\mu^{n-4}}{n - 4}
  \right) \int \exd^n x \sqrt{-g}
 \; {\rm tr} \, a_2\,.
\ee
These can be absorbed into minimal-subtraction (MS) counter-terms
of the form
\be \label{ctform}
 \Sigma_{ct} = \left(
 \frac{1}{4\pi} \right)^{2} \int \exd^4 x \sqrt{-g}
 \; \left[\frac{1}{n - 4} - \frac12
 \ln \left( \frac{m^2}{\mu^2} \right) \right]
 \, {\rm tr} \, a_2 \,,
\ee
and the counter-terms in other renormalization schemes differ from
these by finite renormalizations. Notice that for the counterterm
action unitarity requires that all fields are evaluated at $n =
4$.

Comparing this with the counter-term action and differentiating
with respect to $\mu$ shows (in minimal subtraction) that the
renormalized couplings satisfy the renormalization group equations
\bea
 \mu \frac{\partial m^2}{\partial \mu} =
 \frac{\lambda \, m^2 N}{48 \pi^2} \left( 1 + \frac2N
 \right)\,,\quad && \quad
 \mu \frac{\partial \xi}{\partial \mu} =
 \frac{\lambda \, N}{48 \pi^2} \left(\xi + \frac16
 \right)  \left( 1 + \frac2N
 \right)\nn\\
 \mu \frac{\partial \lambda}{\partial \mu} =
 \frac{\lambda^2 N}{48 \pi^2} \left( 1 + \frac8N \right)
 \,,\quad && \quad
 \mu \frac{\partial V_0}{\partial \mu} =
 \frac{m^4 N}{32 \pi^2} \nn\\
 \mu \frac{\partial A}{\partial \mu} =
 \frac{m^2 N}{16 \pi^2} \left( \xi + \frac16 \right)
 \,,\quad && \quad
 \mu \frac{\partial B}{\partial \mu} =
 -\mu \frac{\partial C}{\partial \mu} =
 \frac{N}{2880 \pi^2} \nn\\
 \mu \frac{\partial D}{\partial \mu} =
 \frac{N}{32 \pi^2} \left( \xi + \frac16 \right)^2
  \,,\quad && \quad
   \mu \frac{\partial E}{\partial \mu} =
   - \frac{N}{96\pi^2} \left( \xi + \frac15 \right)
 \,.
\eea

These may be integrated in the usual way. Write the first few of
these equations as
\be
 \mu \, \frac{\partial \xi}{\partial \mu}
 = b_1\, \lambda \left( \xi + \frac16 \right) \,,
 \quad
  \mu \, \frac{\partial m^2}{\partial \mu}
 = b_1 \, \lambda \, m^2 \,,
\ee
and
\be
 \mu \, \frac{\partial \lambda}{\partial \mu}
 = b_2 \, \lambda^2  \,,
\ee
where
\be
 b_1 = \frac{N}{48 \pi^2} \left(1 + \frac{2}{N} \right)
 \quad \hbox{and} \quad
 b_2 = \frac{N}{48 \pi^2} \left(1 + \frac{8}{N} \right)
 \,.
\ee
The $\lambda$ equation integrates to give
\be
 \lambda(\mu) = \frac{ \lambda_0}{1 - b_2 \,
 \lambda_0 \, \ln(\mu/\mu_0)}
 \,,
\ee
and the other two give
\be
 \frac{\xi(\mu) + 1/6}{\xi_0 + 1/6}
 = \frac{m^2(\mu)}{m^2_0}
 = \left[ \frac{\lambda(\mu)}{\lambda_0}
 \right]^{p} \,,
\ee
where
\be
 p = \frac{b_1}{b_2} = \frac{N+2}{N+8} = 1 - \frac{6}{N}
 + \cdots  \,,
\ee
and so $1 > p \ge \frac13$, with the smallest value occurring when
$N = 1$.

In particular,
\be
 m^2 - 12 \, \xi H^2 =  m^2_0 \left( \frac{\lambda}{\lambda_0}
 \right)^p - 12 \left( \xi_0 +
 \frac16 \right) \left( \frac{\lambda}{\lambda_0}
 \right)^p  + 2 H^2  \,.
\ee
Recalling that both $b_2$ and $p = b_1/b_2$ are positive, we know
$\lambda$ is not asymptotically free, and so $\mu < \mu_0$ implies
both $\lambda < \lambda_0$ and $(\lambda/\lambda_0)^{p} < 1$.
Consequently as $\lambda$ flows to 0 and $\xi$ flows to $-\frac16$
with falling $\mu$, the quantity $m^2 - 12\, \xi H^2$ instead
flows to $2 H^2 > 0$, ensuring $m^2 \to 0$ is necessarily much
smaller than $12\, |\xi| H^2 \to 2H^2$.

The Riemann- and Ricci-squared coefficients are also easily
integrated, giving
\be
 B = B_0 + \frac{N}{2880 \pi^2} \, \ln \left(
 \frac{\mu}{\mu_0} \right)
 \quad \hbox{and} \quad
  C = C_0 -
 \frac{N}{2880 \pi^2} \, \ln \left( \frac{\mu}{\mu_0} \right) \,.
\ee
Those for $A$, $V_0$ and $D$ require integrating the previous
expressions for $m^2(\mu)$ and $\xi(\mu)$, with
\bea
 A(\mu) &=& A_0 + \frac{N}{16\pi^2} \int_{\mu_0}^\mu
 \frac{\exd \mu'}{\mu'} \; m^2(\mu') \left( \xi(\mu')
 + \frac16 \right)  \nn\\
 &=& A_0 + \frac{N  m_0^2}{16\pi^2} \, \left( \xi_0
 + \frac16 \right) \int_{\mu_0}^\mu \frac{\exd \mu'}{\mu'}
 \left[ \frac{\lambda(\mu')}{\lambda_0} \right]^{2p} \nn\\
 &=& A_0 + \frac{N  m_0^2}{16\pi^2} \, \left( \xi_0
 + \frac16 \right) \int_0^{\ln(\mu/\mu_0)} \frac{ \exd t}{
 (1 - b_2 \lambda_0 \, t )^{2p}} \nn\\
 &=& A_0 + \frac{N  m_0^2}{16\pi^2 \lambda_0(2b_1 - b_2)}
 \, \left( \xi_0 + \frac16 \right) \left\{ \frac{1}{
 [1 - b_2 \lambda_0 \, \ln(\mu/\mu_0) ]^{2p-1}} - 1 \right\} \nn\\
 V_0(\mu) &=& V_0(\mu_0) + \frac{N m_0^4}{32\pi^2}
 \int_{\mu_0}^\mu \frac{\exd \mu'}{\mu'}
 \left[ \frac{\lambda(\mu')}{\lambda_0} \right]^{2p} \\
 &=& V_0(\mu_0) + \frac{N  m_0^4}{32\pi^2 \lambda_0(2b_1 - b_2)}
 \left\{ \frac{1}{
 [1 - b_2 \lambda_0 \, \ln(\mu/\mu_0) ]^{2p-1}} - 1 \right\} \nn\\
 D(\mu) &=& D_0 + \frac{N}{32\pi^2} \left( \xi_0 + \frac16
 \right)^2 \int_{\mu_0}^\mu \frac{\exd \mu'}{\mu'} \;
 \left[ \frac{\lambda(\mu')}{\lambda_0} \right]^{2p} \nn\\
 &=& D_0 + \frac{N}{32\pi^2 \lambda_0(2b_1 - b_2)}
 \left( \xi_0 + \frac16 \right)^2 \left\{ \frac{1}{
 [1 - b_2 \lambda_0 \, \ln(\mu/\mu_0) ]^{2p-1}} - 1 \right\}
 \,.\nn
\eea

\subsection*{Running in the large-$N$ limit}

When taking the large-$N$ limit we write $\lambda = g/N$ and may
take $b_1 = b_2 = N b$, up to subdominant terms in $1/N$, with
\be
 b = \frac{1}{48 \pi^2} \,.
\ee
Because $b_1 = b_2$ at large $N$ we also have $p = b_1/b_2 = 1$ in
this limit. The first few RG equations in this limit then become
\be
 \mu \frac{\partial m^2}{\partial \mu} =
 b \, g \, m^2 \,,\qquad
 \mu \frac{\partial \xi}{\partial \mu} =
 b \, g \left(\xi + \frac16  \right)
 \,, \qquad
 \mu \frac{\partial g}{\partial \mu} =
 b \, g^2 \,,
\ee
and so on. These integrate to give
\be
 \frac{\xi(\mu) + 1/6}{\xi_0 + 1/6}
 = \frac{m^2(\mu)}{m^2_0}
 = \frac{g(\mu)}{g_0}
 = \frac{ 1}{1 - b_2 \,
 g_0 \, \ln(\mu/\mu_0)}\,,
\ee
which show that $m^2$, $\xi + \frac16$ and $g$ all renormalize in
the same way, up to $1/N$ corrections. This implies that the
relative size of all of the terms in expressions like
\pref{MlargeN},
\bea
 M^2_\ssN &=& m^2 - 12 \, \xi H^2
 + \frac{g \sigma}{6} \nn\\
 &=& 2H^2 + m^2 - 12 \,
 \left( \xi + \frac16 \right) H^2 + \frac{g \sigma}{6} \,,
\eea
do not change as $\mu$ scales, and $M^2$ and $M^2_\ssN$ both
approach $2H^2$ as $\mu \to 0$.

\end{document}